\RequirePackage{lineno}
\documentclass[aps,twocolumn,showpacs,byrevtex,prl,reprint]{revtex4-1}

\usepackage{graphicx}
\usepackage{dcolumn}
\usepackage{bm}
\usepackage{rotating}
\usepackage{epstopdf}
\usepackage{color}
\usepackage{verbatim} 
\usepackage{multirow}
\usepackage[abs]{overpic}
\usepackage{amsmath}
\usepackage{amssymb}
\usepackage{xspace}

\newcommand{\PreserveBackslash}[1]{\let\temp=\\#1\let\\=\temp}
\newcolumntype{C}[1]{>{\PreserveBackslash\centering}p{#1}}
\newcolumntype{R}[1]{>{\PreserveBackslash\raggedleft}p{#1}}
\newcolumntype{L}[1]{>{\PreserveBackslash\raggedright}p{#1}}

\newcommand{\EE}{e^+e^-}

\newcommand{\too}{\rightarrow}

\begin{document}
\graphicspath{{figure/}}
\DeclareGraphicsExtensions{.eps,.png,.ps}
\title{\boldmath Search for new decay modes of the $\psi_2(3823)$ and the process $e^+e^-\rightarrow\pi^0\pi^0\psi_2(3823)$}
\author{
  \begin{small}
    \begin{center}
      M.~Ablikim$^{1}$, M.~N.~Achasov$^{10,c}$, P.~Adlarson$^{67}$, S. ~Ahmed$^{15}$, M.~Albrecht$^{4}$, R.~Aliberti$^{28}$, A.~Amoroso$^{66A,66C}$, M.~R.~An$^{32}$, Q.~An$^{63,49}$, X.~H.~Bai$^{57}$
% ...
\\
\vspace{0.2cm}
(BESIII Collaboration)\\
\vspace{0.2cm} {\it
$^{1}$ Institute of High Energy Physics, Beijing 100049, People's Republic of China\\
% ...
$^{4}$ Bochum Ruhr-University, D-44780 Bochum, Germany\\
% ...
$^{10}$ G.I. Budker Institute of Nuclear Physics SB RAS (BINP), Novosibirsk 630090, Russia\\
% ...
$^{15}$ Helmholtz Institute Mainz, Johann-Joachim-Becher-Weg 45, D-55099 Mainz, Germany\\
% ...
$^{28}$ Johannes Gutenberg University of Mainz, Johann-Joachim-Becher-Weg 45, D-55099 Mainz, Germany\\
% ...
$^{32}$ Liaoning Normal University, Dalian 116029, People's Republic of China\\
% ...
$^{49}$ State Key Laboratory of Particle Detection and Electronics, Beijing 100049, Hefei 230026, People's Republic of China\\
% ...
$^{57}$ University of Jinan, Jinan 250022, People's Republic of China\\
% ...
$^{63}$ University of Science and Technology of China, Hefei 230026, People's Republic of China\\
% ...
$^{66}$ University of Turin and INFN, (A)University of Turin, I-10125, Turin, Italy; (B)University of Eastern Piedmont, I-15121, Alessandria, Italy; (C)INFN, I-10125, Turin, Italy\\
$^{67}$ Uppsala University, Box 516, SE-75120 Uppsala, Sweden\\
% ...
$^{c}$ Also at the Novosibirsk State University, Novosibirsk, 630090, Russia\\
% ...
      }\end{center}
    \vspace{0.4cm}
\end{small}
}
\affiliation{}


\begin{abstract}
The decays $\psi_2(3823)\rightarrow\gamma\chi_{c0,1,2}, \pi^+\pi^-J/\psi, \pi^0\pi^0J/\psi, \eta J/\psi$, and $\pi^0J/\psi$ are searched for using the reaction $e^+e^-\rightarrow\pi^+\pi^-\psi_2(3823)$ in a 19 fb$^{-1}$ data sample collected at center-of-mass energies between 4.1 and 4.7 GeV with the BESIII detector. The process $\psi_2(3823)\too\gamma\chi_{c1}$ is observed in a 9 fb$^{-1}$ data sample in the center-of-mass energy range 4.3 to 4.7 GeV, which confirms a previous observation but with a higher significance of $11.8\sigma$, and evidence for $\psi_2(3823)\rightarrow\gamma\chi_{c2}$ is found with a significance of $3.2\sigma$ for the first time. The branching-fraction ratio $\frac{\mathcal{B}(\psi_2(3823)\too\gamma\chi_{c2})}{\mathcal{B}(\psi_2(3823)\too\gamma\chi_{c1})}$ is determined. No significant $\psi_2(3823)$ signals are observed for any of the other decay channels. Upper limits of branching-fraction ratios for $\psi_2(3823)\rightarrow\pi^+\pi^-J/\psi, \pi^0\pi^0J/\psi, \eta J/\psi, \pi^0J/\psi, \gamma\chi_{c0}$ relative to $\psi_2(3823)\rightarrow\gamma\chi_{c1}$ are reported. The process $e^+e^-\rightarrow\pi^0\pi^0\psi_2(3823)$ is also searched for, and we find evidence for the process with a significance of $4.3\sigma$. The average cross-section ratio $\frac{\sigma(e^+e^-\rightarrow\pi^0\pi^0\psi_2(3823))}{\sigma(e^+e^-\rightarrow\pi^+\pi^-\psi_2(3823))}$ is also determined.
\end{abstract}


\maketitle

Charmonium, the bound state of a charm quark and anticharm quark ($c\bar{c}$), plays an important role in our understanding of quantum chromodynamics (QCD), which is the fundamental theory of the strong interaction. Low energy QCD remains a field of high interest both experimentally and theoretically. All charmonium states below the open-charm ($D\bar{D}$) threshold have been observed experimentally and can be well described by potential models~\cite{quarkmodel}. However, the understanding of the spectrum that is above the $D\bar{D}$ threshold remains unsettled. During the past decade, many new charmonium-like states have been discovered, such as the $X(3872)$, $Y(4260)$, $Z_{c}(3900)$, and $Z_{cs}(3985)$~\cite{X3872, Y4260, Zc3900, Zcs3985}. These are good candidates for exotic states that lie outside the conventional quark model as discussed in Refs.~\cite{theory11, theory22, theory33, theory44}. On the other hand, there are still excited charmonium states above the $D\bar{D}$ threshold predicted by potential models, which have not yet been observed. Thus, a more complete understanding of the charmonium(-like) spectrum is necessary to identify conventional and exotic states.

The lightest charmonium resonance above the $D\bar{D}$ threshold is the $\psi(3770)$, which is identified as the $\psi(1^{3}D_{1})$ state, the $J=1$ member of the $D$-wave spin-triplet~\cite{pdg}. Recently, two more states have been observed, which are considered to be good candidates for members of this spin-triplet.
The $\psi_2(3823)$, for which first evidence was found by the Belle Collaboration and which was later observed by the BESIII Collaboration in $\psi_2(3823)\too\gamma\chi_{c1}$, is considered to be the $\psi(1^{3}D_{2})$ state~\cite{belle, bes}. The LHCb Collaboration also observed the $\psi_2(3823)$ in its decay to  $\pi^+\pi^-J/\psi$~\cite{lhcb1}. The other newly observed resonance is the $\psi_3(3842)$ seen by LHCb Collaboration in $\psi_3(3842)\too D\bar{D}$~\cite{lhcb2}. It is suggested to be the $\psi(1^{3}D_{3})$ state.

The motivation of this Letter is to provide additional experimental evidence for the correct assignment of the $\psi_{2}(3823)$ to be the $J=2$ spin-triplet partner, by comparing its decay channels to the theory predictions of Refs. ~\cite{theory1, theory2, theory3, theory4, theory5, theory6, theory7, theory8, theory9, theory10}. Experimental information on the $\psi_2(3823)$ is still sparse. The partial widths for decays of the $\psi(1^{3}D_{2})$ state to several channels have been predicted by various different models. These models agree that the dominant decay of the  $\psi(1^{3}D_{2})$  is to $\gamma\chi_{c1}$, with the next most probable decays being to $\gamma\chi_{c2}$ and to  $\pi^+\pi^-J/\psi$. The branching-fraction ratios $\frac{\mathcal{B}(\psi(1^{3}D_{2})\too\gamma\chi_{c2})}{\mathcal{B}(\psi(1^{3}D_{2})\too\gamma\chi_{c1})}$ and $\frac{\mathcal{B}(\psi(1^{3}D_{2})\too\pi^+\pi^-J/\psi)}{\mathcal{B}(\psi(1^{3}D_{2})\too\gamma\chi_{c1})}$ are predicted to be $0.19\sim0.32$ and $0.12\sim0.39$, respectively~\cite{theory1, theory2, theory3, theory4, theory5, theory6, theory7, theory8, theory9, theory10}.

In this Letter, a search for $\psi_2(3823)\too\gamma\chi_{c0,1,2}$, $\pi^+\pi^-J/\psi$, $\pi^0\pi^0J/\psi$, $\eta J/\psi$, and $\pi^0J/\psi$ is reported, using $e^+e^-\rightarrow\pi^+\pi^-\psi_2(3823)$ events from a 19 fb$^{-1}$ data sample collected at center-of-mass energy in the range $4.1<\sqrt{s}<4.7$ GeV with the BESIII detector~\cite{besiii}. Additionally, a search is performed for the process $e^+e^-\too\pi^0\pi^0\psi_2(3823)$ with $\psi_2(3823)\too\gamma\chi_{c1}$.

The BESIII detector is a magnetic spectrometer located at the Beijing Electron
Positron Collider (BEPCII). For more details on the detector or the accelerator, we refer to Refs.~~\cite{besiii,bepcii,whitepaper}. Simulated samples produced with the {\sc{Geant4}}-based~\cite{geant4} Monte Carlo (MC) package, which
includes the geometric description of the BESIII detector and the
detector response, are used to determine the detection efficiency
and to estimate background contributions. The simulation includes the beam
energy spread and initial-state radiation (ISR) in $e^+e^-$
annihilations modeled with the generator {\sc kkmc}~\cite{KKMC}.
Signal MC samples for $\EE \too \pi\pi \psi_2(3823)$ are generated using isotropic phase space populations, assuming that the cross section follows a coherent sum of $\psi(4360)$ and the $\psi(4660)$ Breit-Wigner (BW) distributions, whose magnitude and phase parameters we obtain from a fit to the observed cross section, with the $\psi_2(3823)$ mass fixed to the Particle Data Group (PDG) value~\cite{pdg} and width fixed to zero. The subsequent $\psi_2(3823)$ decays are generated uniformly in the phase space, and the effects from the angular distributions of $\psi_2(3823)$ decays are studied and found to be small.
Inclusive MC samples consist of the production of open charm
processes, the ISR production of vector charmonium(-like) states,
and the continuum processes incorporated in {\sc kkmc}~\cite{KKMC}. Known decay modes are modeled with {\sc evtgen}~\cite{ref:evtgen} using branching fractions summarized and averaged by the
PDG~\cite{pdg}. The remaining unknown decays
from the charmonium states are generated with {\sc lundcharm}~\cite{ref:lundcharm}. Final state radiation from charged final-state particles is incorporated with the {\sc photos} package~\cite{photos}.

The $\chi_{c1,2}$ are reconstructed via $\chi_{c1,2}\too\gamma J/\psi$ decays, the $J/\psi$ is reconstructed in its decay to an $e^+e^-$ or $\mu^+\mu^-$ pair, the $\pi^0$ and $\eta$ are reconstructed via $\pi^0/\eta\too\gamma\gamma$ decays, and the $\chi_{c0}$ is reconstructed in its decay to a $\pi^+\pi^-$ or $K^+K^-$ pair. For each charged track, the distance of closest approach to the interaction point is required to be within $\pm10$ cm in the beam direction and within 1 cm in the plane perpendicular to the beam direction. The polar angle ($\theta$) of the tracks must be within the fiducial volume of the multilayer drift chamber $(|\cos\theta|<0.93)$. Photons are reconstructed from isolated showers in the electromagnetic calorimeter (EMC), which are at least $10^\circ$ away from the nearest charged track. The photon energy is required to be at least 25 MeV in the barrel region $(|\cos\theta|<0.8)$ or 50 MeV in the end-cap region $(0.86<|\cos\theta|<0.92)$. To suppress electronic noise and energy depositions unrelated to the event, the time at which the photon is recorded in the EMC is required to be within $700$ ns of the event start time. Candidate events must have the exact same number of charged tracks with zero net charge and at least the same number of photons as required for the respective final state. Tracks with momenta larger than 1~GeV/$c$ are assigned to be  leptons from the decay of a $J/\psi$ or to be $\pi/K$ from the decay of a $\chi_{c0}$. Otherwise, tracks are considered pions. Leptons from the $J/\psi$ decay with an energy deposit in the EMC larger than 1.0~GeV are identified as electrons, and those with less than 0.4~GeV as muons. To reduce background contributions and to improve the mass resolution, a four-constraint kinematic fit is performed to constrain the total four-momentum of the final state particles to the four-momentum of the colliding beams. Additionally, for the $\psi_2(3823)\too\pi^0\pi^0J/\psi$ and $e^+e^-\too\pi^0\pi^0\psi_2(3823)$ channels the invariant masses of the two pairs of photons are constrained to the nominal mass of the $\pi^0$ meson~\cite{pdg}. The two track candidates from the decay of $\chi_{c0}$ mesons are considered to be either a $\pi^{+}\pi^{-}$ or a $K^{+}K^{-}$ pair depending on the $\chi^2$ of the four-constraint kinematic fit. If $\chi^{2}(\pi^{+}\pi^{-}) < \chi^{2}(K^{+}K^{-})$, the two tracks are identified as a $\pi^{+}\pi^{-}$ pair, otherwise, as a $K^{+}K^{-}$ pair. For all these channels, if there is more than one combination of photons in an event, the one with the smallest $\chi^{2}$ of the kinematic fit is selected. The $\chi^{2}$ of the candidate process is required to be less than 60 in all cases.

Besides the requirements described above, further selection criteria are applied. To suppress the background from $\pi^0/\eta\too\gamma\gamma$ in $\psi_2(3823)\too\gamma\chi_{c1,2}$ decays,  regions around the $\pi^0$ and $\eta$ masses, namely $[0.11, 0.16]$~GeV/$c^2$ and $[0.51, 0.58]$~GeV/$c^2$, in the invariant mass $M(\gamma\gamma)$ are excluded. In order to remove background from $\psi(3686)\too\pi^+\pi^-J/\psi$ in $\psi_2(3823)\too\pi^+\pi^-J/\psi, \pi^0\pi^0J/\psi, \eta J/\psi, \pi^0J/\psi$ decays, all possible invariant mass $M(\pi^+\pi^-J/\psi)$ combinations are required to be outside the region $[3.675, 3.696]$~GeV/$c^2$. To eliminate background from $\eta'\too\eta\pi^+\pi^-$ and $\chi_{c1}\too\gamma J/\psi$ in $\psi_2(3823)\too\eta J/\psi$ decays, the invariant masses $M(\gamma\gamma\pi^+\pi^-)$ and $M(\gamma_{H}J/\psi)$ are required to be outside the regions $[0.94, 0.97]$~GeV/$c^2$ and $[3.49, 3.53]$~GeV/$c^2$ respectively, where $\gamma_{H}$ is the highest energy photon. This condition removes almost all of the $\eta'/\chi_{c1}$ background. To remove background from $\eta\too\pi^0\pi^+\pi^-$ in $\psi_2(3823)\too\pi^0J/\psi$ decays, candidates are excluded that have an  invariant mass $M(\gamma\gamma\pi^+\pi^-)$ around the nominal $\eta$ mass in the region $[0.51, 0.58]$~GeV/$c^2$. Finally, to reject  background from photon conversion in $\psi_2(3823)\too\gamma\chi_{c0}$ decays, the cosine of the angle between any two charged tracks is required to be less than 0.9.

The $J/\psi$ signal region is defined by the mass range [3.075, 3.125]~GeV/$c^{2}$ in $M(\EE/\mu^+\mu^-)$, apart from in the decay channel $\psi_2(3823)\too\pi^+\pi^-J/\psi$ where the $J/\psi$ signal region is narrowed to the range [3.09, 3.11]~GeV/$c^{2}$ due to the better resolution for the four charged-track final states. The $\chi_{c1}$ and $\chi_{c2}$ signal regions are chosen as the ranges $[3.49, 3.53]~\text{GeV}/c^{2}$ and $[3.54, 3.57]~\text{GeV}/c^{2}$ in $M(\gamma_{H}J/\psi)$, respectively, and sideband regions, defined as the ranges $[3.43, 3.48]~\text{GeV}/c^{2}$ and $[3.58, 3.63]~\text{GeV}/c^{2}$, are used to study the non-resonant background. The $\eta$, $\pi^0$ and $\chi_{c0}$ signal regions are chosen to be [0.52, 0.57] GeV$/c^{2}$ and [0.12, 0.15] GeV$/c^{2}$ in $M(\gamma\gamma)$, and $[3.39, 3.44]~\text{GeV}/c^{2}$ in $M(\pi^+\pi^-/K^+K^-)$, respectively.

\begin{figure}[htbp]
\begin{center}
\begin{overpic}[width=0.23\textwidth]{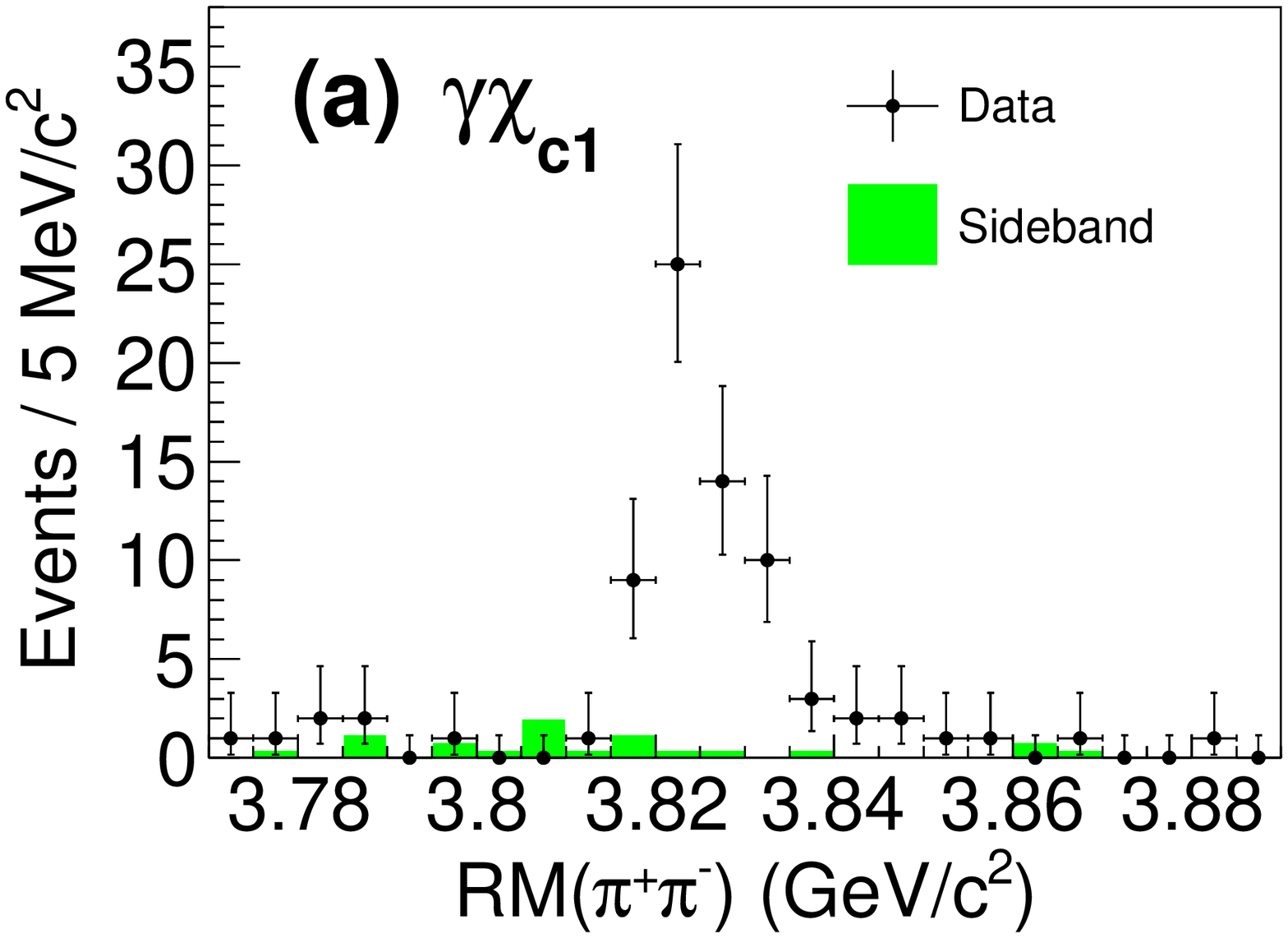}
\end{overpic}
\begin{overpic}[width=0.23\textwidth]{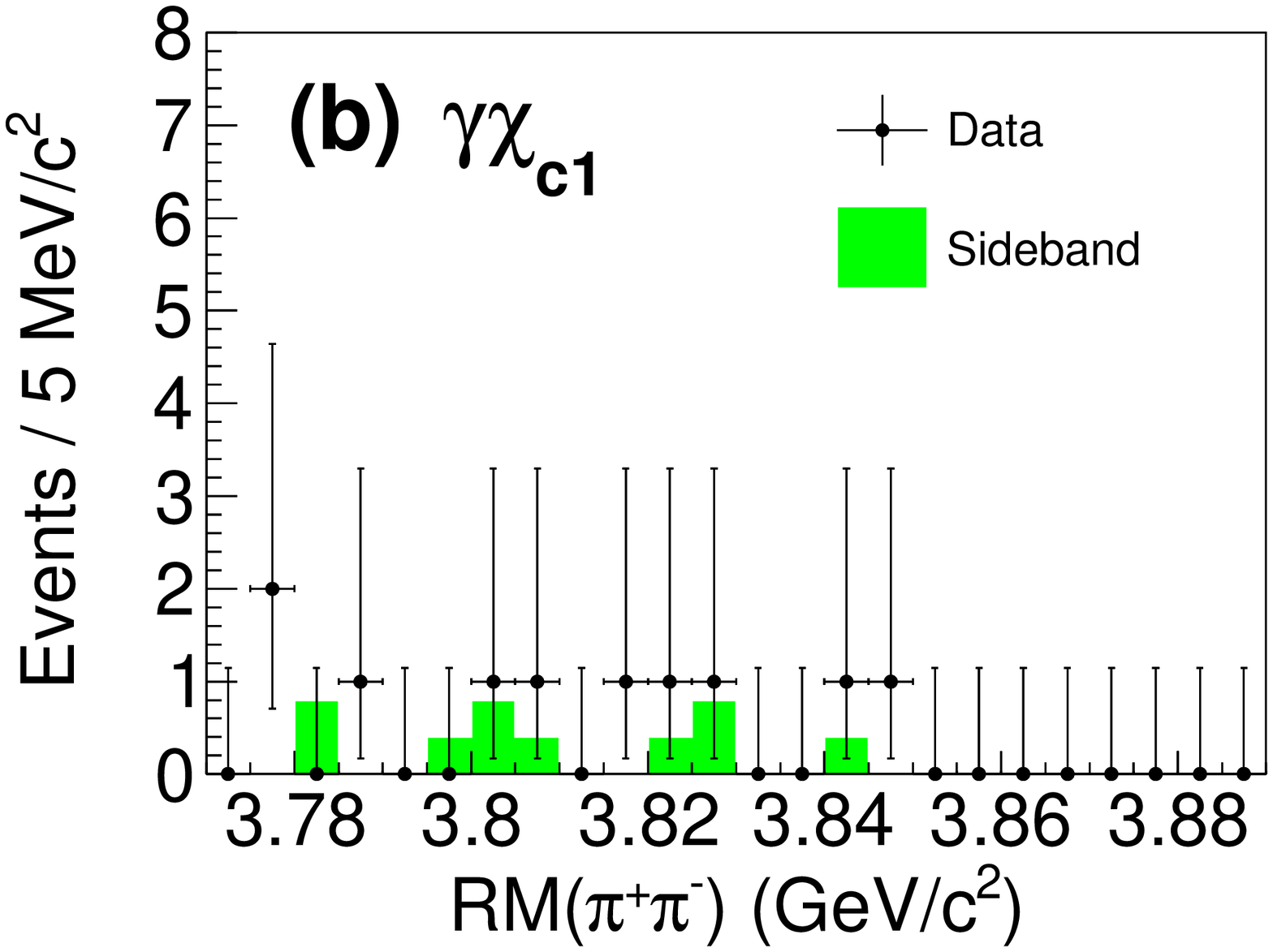}
\end{overpic}
\caption{$\pi^+\pi^-$ recoil-mass distribution $RM(\pi^+\pi^-)$ for $\gamma\chi_{c1}$ channel for the data at $4.3<\sqrt{s}<4.7$ GeV (a) and data at $4.1<\sqrt{s}<4.3$ GeV (b). The green shaded histograms correspond to the normalized events from the $\chi_{c1}$ sideband region.}
\label{fig:m145}
\end{center}
\end{figure}

Figure~\ref{fig:m145} shows the $\pi^+\pi^-$ recoil-mass distribution $RM(\pi^+\pi^-)$ for the $\gamma\chi_{c1}$ channel.
A clear $\psi_2(3823)$ signal is observed for 9~fb$^{-1}$ of data at $4.3<\sqrt{s}<4.7$~GeV, while no significant $\psi_2(3823)$ signal is seen for the 10~fb$^{-1}$ sample at $4.1<\sqrt{s}<4.3$ GeV. The green shaded histograms correspond to the normalized events from the $\chi_{c1}$ sideband region. Thus, only data at $4.3<\sqrt{s}<4.7$~GeV are used to search for new $\psi_2(3823)$ decay channels.

\begin{figure}[htbp]
\begin{center}
\begin{overpic}[width=0.48\textwidth]{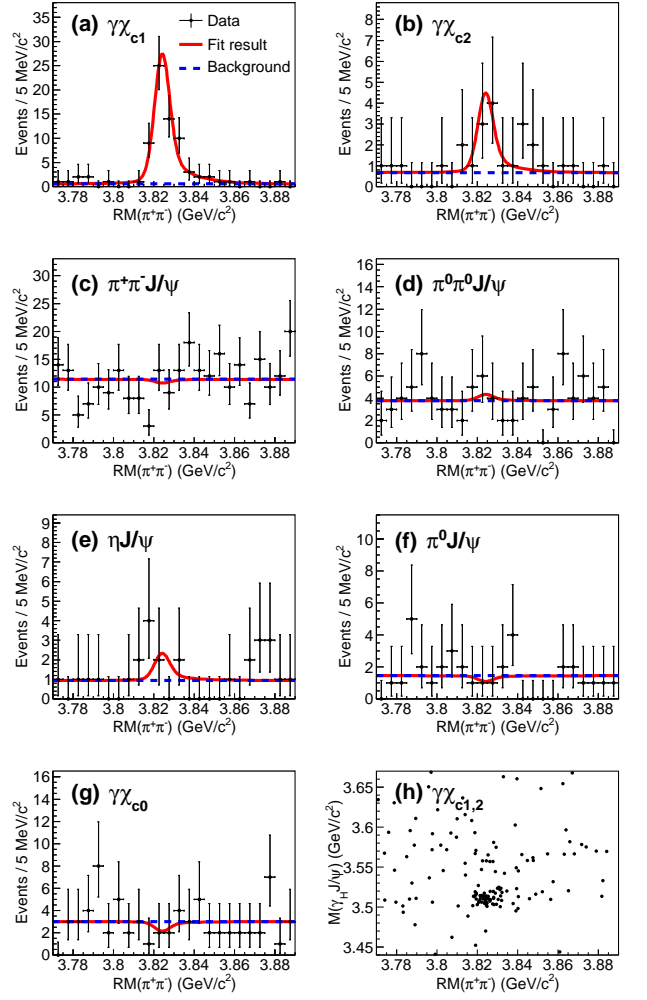}
\end{overpic}
\caption{Results of the simultaneous fits to the seven distributions of $RM(\pi^+\pi^-)$ for the decays $\psi_2(3823)\too\gamma\chi_{c1}$ (a), $\gamma\chi_{c2}$ (b), $\pi^+\pi^-J/\psi$ (c), $\pi^0\pi^0J/\psi$ (d), $\eta J/\psi$ (e), $\pi^0J/\psi$ (f), $\gamma\chi_{c0}$ (g), and a scatter plot of $M(\gamma_{H}J/\psi)$ versus $RM(\pi^+\pi^-)$ for the decays $\psi_2(3823)\too\gamma\chi_{c1,2}$ (h) for data at $4.3<\sqrt{s}<4.7~\text{GeV}$.
The red solid lines are the total fit results and the blue dashed lines are the background components.}
\label{fig:fit}
\end{center}
\end{figure}

Figure~\ref{fig:fit} shows the distributions of $RM(\pi^+\pi^-)$ for the decays $\psi_2(3823)\too\gamma\chi_{c1}$, $\gamma\chi_{c2}$, $\pi^+\pi^-J/\psi$, $\pi^0\pi^0J/\psi$, $\eta J/\psi$, $\pi^0J/\psi$, $\gamma\chi_{c0}$ and a scatter plot of $M(\gamma_{H}J/\psi)$ versus $RM(\pi^+\pi^-)$ for the decays $\psi_2(3823)\too\gamma\chi_{c1,2}$ for data at $4.3<\sqrt{s}<4.7~\text{GeV}$. Here, all valid $RM(\pi^+\pi^-)$ combinations of the $\pi^+\pi^-J/\psi$ decay are retained. In addition to the $\psi_2(3823)$ signal observed in the $\psi_2(3823)\too\gamma\chi_{c1}$ channel, there are also events clustered in the signal region for the mode $\psi_2(3823)\too\gamma\chi_{c2}$. No significant $\psi_2(3823)$ signals are observed for the other channels. The distribution of $M(\pi^+\pi^-J/\psi)$ after a four-constraint kinematic fit for the $\pi^+\pi^-J/\psi$ decay is also checked, but no significant $\psi_2(3823)$ signals are seen. Furthermore, in any of these channels, no significant $e^+e^-\rightarrow\pi^+\pi^-\psi_3(3842)$ signals are found. A detailed study of the inclusive MC samples indicates that there are no peaking background contributions in the $\psi_2(3823)$ signal region~\cite{inclusive}. In order to extract the $\psi_2(3823)$ signal yield, a simultaneous unbinned maximum-likelihood fit is performed to the seven decay channels. The expected shape of $RM(\pi^+\pi^-)$ from the signal process is modeled by the shape from the MC simulation convolved with a Gaussian function. The parameters of mean and width are free parameters in the fit, but are constrained to be the same in all channels. The background is described by a constant. The solid curves in Fig.~\ref{fig:fit} show the fit results. The significances with systematic uncertainty included for the decays $\psi_2(3823)\too\gamma\chi_{c1}$ and $\psi_2(3823)\too\gamma\chi_{c2}$ are $11.8\sigma$ and $3.2\sigma$, respectively. For the other decays, where there are no significant signals, upper limits of the relative branching ratio compared to the decay $\psi_2(3823)\too\gamma\chi_{c1}$ at the $90\%$ confidence level (C.L.) are determined. These upper limits are calculated from the likelihood curve of the fits as a function of signal yield after being convolved with a Gaussian distribution, where the width of Gaussian distribution is the quadratic sum of the systematic uncertainty and statistical uncertainty of the $\psi_2(3823)\too\gamma\chi_{c1}$ signal yield. Those limits together with the corresponding limits on the number of signal events are summarized in Table~\ref{tab:fitresult}.

\begin{table}[htbp]
\begin{center}
\caption{ The number of $\psi_2(3823)$ signal events $N^{\psi_2(3823)}$ and branching-fraction ratios $\frac{\mathcal{B}(\psi_2(3823)\too ...)}{\mathcal{B}(\psi_2(3823)\too\gamma\chi_{c1})}$ for different $\psi_2(3823)$ decay channels. For $N^{\psi_2(3823)}$ only the statistical uncertainty is shown. For the ratios the first uncertainty is statistical and the second uncertainty is systematic. The upper limits at the $90\%$ C.L. are calculated taking into account both contributions. Dash means that the result is not applicable.}
\label{tab:fitresult}
\begin{tabular}{ccc}
  \hline
  \hline
  \ \ \ \ \ Channel \ \ \ \ \ & \ \ \ \ \ $N^{\psi_2(3823)}$ \ \ \ \ \ & \ \ \ \ \ $\frac{\mathcal{B}(\psi_2(3823)\too ...)}{\mathcal{B}(\psi_2(3823)\too\gamma\chi_{c1})}$ \ \ \ \ \ \\
  \hline
  $\gamma\chi_{c1}$ & $63.1\pm8.5$ & $-$ \\
  $\gamma\chi_{c2}$ & $8.8^{+4.3}_{-3.4}$ & $0.28^{+0.14}_{-0.11}\pm0.02$ \\
  $\pi^+\pi^-J/\psi$ & $<21.0$ & $<0.06$ \\
  $\pi^0\pi^0J/\psi$ & $<10.0$ & $<0.11$ \\
  $\eta J/\psi$ & $<9.8$ & $<0.14$ \\
  $\pi^0J/\psi$ & $<5.6$ & $<0.03$ \\
  $\gamma\chi_{c0}$ & $<6.3$ & $<0.24$ \\
  \hline
  \hline
\end{tabular}
\end{center}
\end{table}

\begin{table*}[htbp]
\caption{Definitions of the ratios $\frac{\mathcal{B}(\psi_2(3823)\too...)}{\mathcal{B}(\psi_2(3823)\too\gamma\chi_{c1})}$ and $\frac{\sigma(\EE\too\pi^0\pi^0\psi_2(3823))}{\sigma(\EE\too\pi^+\pi^-\psi_2(3823))}$, where $\mathcal{B}(\psi_2(3823)\too...)$ represents the branching fraction of $\psi_2(3823)$ decays into a certain channel, and $\mathcal{B}(...)$ represents the branching fraction of subsequent decays.}
\label{tab:definition}
\begin{tabular}{c c}
  \hline
  \hline
  Ratio & Definition \\
  \hline
  \ \ \ \ \ \ \ \ \large{$\frac{\mathcal{B}(\psi_2(3823)\too...)}{\mathcal{B}(\psi_2(3823)\too\gamma\chi_{c1})}$} \ \ \ \ \ \ \ \ & \ \ \ \ \ \ \ \ \large{$\frac{N^{\psi_2(3823)\too...}}{N^{\psi_2(3823)\too\gamma\chi_{c1}}}
    \frac{\sum\limits_i\mathcal{L}_i\sigma_i(1+\delta)_i\epsilon_i^{\psi_2(3823)\too\gamma\chi_{c1}}}{\sum\limits_i\mathcal{L}_i\sigma_i(1+\delta)_i\epsilon_i^{\psi_2(3823)\too...}}
    \frac{\mathcal{B}(\chi_{c1}\too\gamma J/\psi\too\gamma l^+l^-)}{\mathcal{B}(...)}$} \ \ \ \ \ \ \ \  \\
  \large{$\frac{\sigma(\EE\too\pi^0\pi^0\psi_2(3823))}{\sigma(\EE\too\pi^+\pi^-\psi_2(3823))}$} & \large{$\frac{N^{\pi^0\pi^0\psi_2(3823)}}{N^{\pi^+\pi^-\psi_2(3823)}}
    \frac{\sum\limits_i\mathcal{L}_i(1+\delta)_i\epsilon_i^{\pi^+\pi^-\psi_2(3823)}}{\sum\limits_i\mathcal{L}_i(1+\delta)_i\epsilon_i^{\pi^0\pi^0\psi_2(3823)}}
    \frac{1}{\mathcal{B}^{2}(\pi^0\too\gamma\gamma)}$}  \\
  \hline
  \hline
\end{tabular}
\end{table*}

The values of the branching-fraction ratio $\frac{\mathcal{B}(\psi_2(3823)\too\gamma\chi_{c2})}{\mathcal{B}(\psi_2(3823)\too\gamma\chi_{c1})}$ and the upper limits of the branching-fraction ratios for $\psi_2(3823)\rightarrow \pi^+\pi^-J/\psi$, $\pi^0\pi^0J/\psi$, $\eta J/\psi$, $\pi^0J/\psi$ and $\gamma\chi_{c0}$ relative to the decay $\psi_2(3823)\rightarrow\gamma\chi_{c1}$ shown in Table~\ref{tab:fitresult} are calculated using the definition in Table~\ref{tab:definition}, where $N$ is the yield of signal events, $\mathcal{L}$ is the integrated luminosity~\cite{luminosity}, $\sigma$ is the cross section, $1+\delta$ is the radiative correction factor~\cite{KKMC, QED}, $\epsilon$ is the efficiency, $\mathcal{B}$ is the branching fraction~\cite{pdg}, and $i$ denotes each energy point.

\begin{figure}[htbp]
\begin{center}
\begin{overpic}[width=0.36\textwidth]{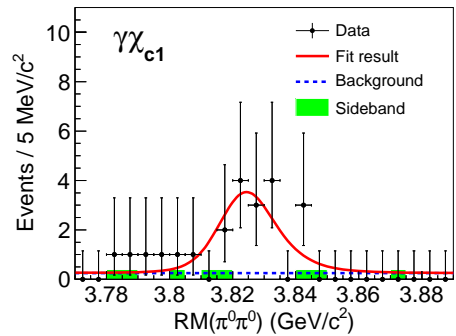}
\end{overpic}
\caption{Results of the fit to the invariant-mass distribution of $RM(\pi^0\pi^0)$ for decay $\psi_2(3823)\too\gamma\chi_{c1}$ for data at $4.3<\sqrt{s}<4.7$ GeV. The red solid line is the total fit to data, the blue dashed line is the background component, and the green shaded histogram corresponds to the normalized events from the $\chi_{c1}$ sideband region.}
\label{fig:fitpi0pi0}
\end{center}
\end{figure}

Figure~\ref{fig:fitpi0pi0} shows the $\pi^0\pi^0$ recoil-mass distribution $RM(\pi^0\pi^0)$ for the decay $\psi_2(3823)\too\gamma\chi_{c1}$ for data at $4.3<\sqrt{s}<4.7$ GeV. A signal peak corresponding to the process  $\EE\too\pi^0\pi^0\psi_2(3823)$ can be seen. In order to determine the signal yield, an unbinned maximum-likelihood fit is performed. The $\psi_2(3823)$ signal is modeled by the MC-determined shape convolved with a Gaussian function, whose mean value and width are fixed to be the values obtained from the same final-state process $\EE\too\pi^0\pi^0\psi(3686)$. The background is described with a constant. The solid curve in Fig.~\ref{fig:fitpi0pi0} shows the fit results. The number of signal events is determined to be $15.9^{+5.1}_{-4.4}$, and the significance for the process $\EE\too\pi^0\pi^0\psi_2(3823)$ with systematic uncertainties included is found to be $4.3\sigma$. The average cross-section ratio $\frac{\sigma(e^+e^-\rightarrow\pi^0\pi^0\psi_2(3823))}{\sigma(e^+e^-\rightarrow\pi^+\pi^-\psi_2(3823))}$ for the $\gamma\chi_{c1}$ channel for data at $4.3<\sqrt{s}<4.7$ GeV is determined to be $0.64^{+0.22}_{-0.20}\pm0.05$, which is calculated using the definition in Table~\ref{tab:definition}, and is consistent with the expectation of isospin symmetry.

\begin{table*}[htbp]
\caption{Relative systematic uncertainties (in $\%$) from the different sources for average cross-section ratio $\frac{\sigma(e^+e^-\rightarrow\pi^0\pi^0\psi_2(3823))}{\sigma(e^+e^-\rightarrow\pi^+\pi^-\psi_2(3823))}$ (first column) and branching-fraction ratios $\frac{\mathcal{B}(\psi_2(3823)\too ...)}{\mathcal{B}(\psi_2(3823)\too\gamma\chi_{c1})}$ (second to sixth columns) for each decay channel. Dashes mean that the results are not applicable or cancel in the ratio.}
\label{tab:sumerror}
\begin{tabular}{c | c | c c c c c c c c c c c c c}
  \hline
  \hline
  \ \ \ \ \ Source \ \ \ \ \ & \ \ \ \ \ $\gamma\chi_{c1}$ \ \ \ \ \ & \ \ \ \ \ $\gamma\chi_{c2}$ \ \ \ \ \ & \ \ \ \ \ $\pi^+\pi^-J/\psi$ \ \ \ \ \ & \ \ \ \ \ $\pi^0\pi^0J/\psi$ \ \ \ \ \ & \ \ \ \ \ $\eta J/\psi$ \ \ \ \ \ & \ \ \ \ \ $\pi^0J/\psi$ \ \ \ \ \ & \ \ \ \ \ $\gamma\chi_{c0}$ \ \ \ \ \ \\
  \hline
  Tracking efficiency & 2.0 & $-$ & 2.0 & $-$ & $-$ & $-$ & $-$  \\
  Photon efficiency   & 4.0 & $-$ & 2.0 & 2.0 & $-$ & $-$ & 1.0  \\
  Branching fraction  & 0.1 & 3.9 & 2.9 & 2.9 & 3.0 & 2.9 & 4.4  \\
  Kinematic fit       & 0.9 & 0.5 & 0.6 & 0.3 & 0.2 & 0.4 & 0.4  \\
  Input line shape    & 0.1 & 0.4 & 1.3 & 1.6 & 1.8 & 0.6 & 0.3  \\
  MC decay model      & 5.8 & 0.7 & 0.8 & 1.0 & 1.3 & 2.1 & 10.5  \\
  Fit method          & 4.2 & 4.5 & $-$ & $-$ & $-$ & $-$ & $-$  \\
  Mass window         & $-$ & $-$ & 1.0 & 1.0 & 1.4 & 1.4 & 2.5  \\
  sum                 & 8.5 & 6.0 & 4.5 & 4.1 & 4.0 & 3.9 & 11.7 \\
  \hline
  \hline
\end{tabular}
\end{table*}

The considered sources of systematic uncertainties related to the branching-fraction ratios and average cross-section ratio are summarized in Table~\ref{tab:sumerror}, where those that are common to the numerator and denominator cancel. The uncertainty in the tracking efficiency and photon efficiency is $1\%$ per track or per photon~\cite{track}. The uncertainty from the branching fractions is taken from the PDG~\cite{pdg}. The uncertainty due to the kinematic fit is estimated by correcting the helix parameters of charged tracks, and the difference between the results with and without this correction is taken as the uncertainty~\cite{helix}. To estimate the uncertainty related to the input line-shape of the process $\EE\too\pi\pi \psi_2(3823)$, we change the input line-shape to a coherent sum of BW functions of $\psi(4415)$ and $\psi(4660)$ with the parameters fixed to PDG values, where magnitude and phase parameters are obtained from a fit to the cross section of $\EE\too\pi^{+}\pi^{-} \psi_2(3823)$. The process $\EE\too\pi\pi \psi_2(3823)$ is generated by the three-body phase-space model, the uncertainty of the MC decay model is obtained by changing the phase-space model to the model $\EE\too f_0(500)\psi_2(3823)$ with a $D$-wave in the MC simulation. The  angular distribution of the angle between the two low-momentum pions in the lab frame is sensitive to the MC model for the $\psi_2(3823)\too\gamma\chi_{c0}$ decay, which leads to the dominant systematic uncertainty contribution for this mode. The uncertainty from the fit range is obtained by varying the limits of the fit range by $\pm5~\text{MeV}/c^{2}$, and the uncertainty associated with the background shape is estimated by changing the constant background to a linear background.  The influence from the possible presence of a $\psi_3(3842)$ state is accounted for by including this component in the fit. In each case, the difference to the nominal result is taken as the systematic uncertainty. The uncertainties from the $J/\psi$, $\pi^0/\eta$, $\chi_{c1,2}$ and $\chi_{c0}$ mass-window requirements are $1.6\%$, $1.0\%$, $1.0\%$, and $1.7\%$, respectively~\cite{jpsimasswindow, pi0masswindow}. The overall systematic uncertainties are obtained by adding all the sources of systematic uncertainties in quadrature, assuming they are uncorrelated. The effect of the systematic uncertainties on the upper limit or significance is accounted for by changing the fit range and the background shape and then choosing the largest value of the upper limit or the lowest value of the significance.

In summary, the decays $\psi_2(3823)\rightarrow\gamma\chi_{c0,1,2}$, $\pi^+\pi^-J/\psi, \pi^0\pi^0J/\psi, \eta J/\psi$, and $\pi^0J/\psi$ are searched for using the process $e^+e^-\rightarrow\pi^+\pi^-\psi_2(3823)$ in a 19 fb$^{-1}$ data sample collected at center-of-mass energy between 4.1 and $4.7~\text{GeV}$ with the BESIII detector. The process $\psi_2(3823)\too\gamma\chi_{c1}$ is observed in a 9 fb$^{-1}$ data sample in the center-of-mass energy range 4.3 to $4.7~\text{GeV}$, which confirms the previous observation but with the higher significance of $11.8\sigma$, and evidence for the process $\psi_2(3823)\rightarrow\gamma\chi_{c2}$ is found for the first time with a significance of $3.2\sigma$. The branching-fraction ratio $\frac{\mathcal{B}(\psi_2(3823)\too\gamma\chi_{c2})}{\mathcal{B}(\psi_2(3823)\too\gamma\chi_{c1})}$ is measured to be $0.28^{+0.14}_{-0.11}\pm0.02$, which is consistent with the theoretical predictions for $\frac{\mathcal{B}(\psi(1^{3}D_{2})\too\gamma\chi_{c2})}{\mathcal{B}(\psi(1^{3}D_{2})\too\gamma\chi_{c1})}$~\cite{theory1, theory2, theory3, theory4, theory5, theory6, theory7, theory8, theory9, theory10}. No significant $\psi_2(3823)$ signals are observed for other channels. The upper limits of branching-fraction ratios for $\psi_2(3823)\rightarrow\pi^+\pi^-J/\psi, \pi^0\pi^0J/\psi, \eta J/\psi, \pi^0J/\psi$ and $\gamma\chi_{c0}$ relative to $\psi_2(3823)\rightarrow\gamma\chi_{c1}$ are reported, and the results can be found in Table~\ref{tab:fitresult}. The upper limit at the $90\%$ C.L. of the branching-fraction ratio $\frac{\mathcal{B}(\psi_2(3823)\too\pi^+\pi^-J/\psi)}{\mathcal{B}(\psi_2(3823)\too\gamma\chi_{c1})}$ is determined to be 0.06, which is lower than the theoretical predictions given in Refs.~\cite{theory1, theory2, theory3, theory4, theory5, theory6, theory7, theory8, theory9, theory10}.
No significant $e^+e^-\rightarrow\pi^+\pi^-\psi_3(3842)$ signals are seen in any of the channels we studied.
The process $e^+e^-\rightarrow\pi^0\pi^0\psi_2(3823)$ with $\psi_2(3823)\too\gamma\chi_{c1}$ is also searched for, and evidence for the process is found with a significance of $4.3\sigma$. The average cross-section ratio $\frac{\sigma(e^+e^-\rightarrow\pi^0\pi^0\psi_2(3823))}{\sigma(e^+e^-\rightarrow\pi^+\pi^-\psi_2(3823))}$ is determined to be $0.64^{+0.22}_{-0.20}\pm0.05$, which is consistent with the expectation of isospin symmetry.

The BESIII collaboration thanks the staff of BEPCII and the IHEP computing center for their strong support. This work is supported in part by National Key Research and Development Program of China under Contracts Nos. 2020YFA0406300, 2020YFA0406400; National Natural Science Foundation of China (NSFC) under Contracts Nos. 11905179, 11625523, 11635010, 11735014, 11822506, 11835012, 11935015, 11935016, 11935018, 11961141012; the Chinese Academy of Sciences (CAS) Large-Scale Scientific Facility Program; Joint Large-Scale Scientific Facility Funds of the NSFC and CAS under Contracts Nos. U1732263, U1832207; CAS Key Research Program of Frontier Sciences under Contracts Nos. QYZDJ-SSW-SLH003, QYZDJ-SSW-SLH040; 100 Talents Program of CAS; INPAC and Shanghai Key Laboratory for Particle Physics and Cosmology; ERC under Contract No. 758462; European Union Horizon 2020 research and innovation programme under Contract No. Marie Sklodowska-Curie grant agreement No 894790; Nanhu Scholars Program for Young Scholars of Xinyang Normal University; German Research Foundation DFG under Contracts Nos. 443159800, Collaborative Research Center CRC 1044, FOR 2359, FOR 2359, GRK 214; Istituto Nazionale di Fisica Nucleare, Italy; Ministry of Development of Turkey under Contract No. DPT2006K-120470; National Science and Technology fund; Olle Engkvist Foundation under Contract No. 200-0605; STFC (United Kingdom); The Knut and Alice Wallenberg Foundation (Sweden) under Contract No. 2016.0157; The Royal Society, UK under Contracts Nos. DH140054, DH160214; The Swedish Research Council; U. S. Department of Energy under Contracts Nos. DE-FG02-05ER41374, DE-SC-0012069.

\end{document}